\documentclass[namedreferences]{SolarPhysics}
\usepackage[optionalrh]{spr-sola-addons} 
\usepackage{epsfig}                     
\usepackage{graphicx}                    
\usepackage{color}                       
\usepackage{url}                         

\usepackage{solaheader}	
\newcommand{\solareceiveddate}{1 April 2009} 
\newcommand{\solaaccepteddate}{2 April 2009}
\newcommand{\solaonlinedate}{3 April 2009}
\newcommand{\doi}{10.1007/s11207-010-9679-0}
\renewcommand{\solavolume}{123}
\renewcommand{\firstpage}{1}
\renewcommand{\lastpage}{8}
\renewcommand{\solayear}{2008}
\begin{document}
\begin{article}
\begin{opening}

\title{Numerical MHD Simulations of Solar Magnetoconvection and Oscillations in Inclined Magnetic Field Regions}
\author{I.N.~\surname{Kitiashvili}$^{1,2}$\sep
        A.G.~\surname{Kosovichev}$^{2}$\sep
        N.N.~\surname{Mansour}$^{3}$\sep
        A.A.~\surname{Wray}$^{3}$}
%
\runningauthor{I.N. Kitiashvili {\it et al.}}
\runningtitle{Numerical Simulations of Solar Magnetoconvection and Oscillations}

%
  \institute{$^{1}$ Center for Turbulence Research, Stanford University, Stanford, CA 94305, USA
                     email: \url{irinasun@stanford.edu}\\
             $^{2}$ Hansen Experimental Physics Laboratory, Stanford University,
Stanford, CA 94305, USA email: \url{sasha@sun.stanford.edu} \\
             $^{3}$ NASA Ames Research Center, Moffett Field, CA 94035, USA \\}

\begin{abstract}
The sunspot penumbra is a transition zone between the strong vertical magnetic field area
(sunspot umbra) and the quiet Sun. The penumbra has a fine filamentary structure that is
characterized by magnetic field lines inclined toward the surface. Numerical simulations
of solar convection in inclined magnetic field regions have provided an explanation of
the filamentary structure and the Evershed outflow in the penumbra. In this paper, we use
radiative MHD simulations to investigate the influence of the magnetic field inclination
on the power spectrum of vertical velocity oscillations. The results reveal a strong shift
of the resonance mode peaks to higher frequencies in the case of a highly inclined magnetic
field. The frequency shift for the inclined field is significantly greater than that in vertical field
regions of similar strength. This is consistent with the behavior of fast MHD waves.
\end{abstract}

%
\end{opening}

\section{Introduction}

Excitation and properties of solar oscillations have been investigated using three-dimensional (3D) numerical
simulations by many authors \cite{nordlund2001,stein2001,georgobiani2003,stein2004,jacoutot08a}.
The results have provided important insights into the excitation mechanism \cite{stein2001} and the meaning of the line asymmetry \cite{georgobiani2003}, and have been used for testing the
local correlation tracking technique \cite{georgobiani2007} and in time-distance helioseismology
\cite{zhao2007}. However, properties of oscillations in magnetic regions using realistic
simulations of solar magnetoconvection have been much less investigated. Analyzing the magnetohydrodynamic (MHD) simulations
with initially vertical magnetic field,  Jacoutot {\it et al.} (2008b, 2009) found a shift
of the oscillation power toward higher frequencies with the increasing magnetic field
strength. They also found enhanced excitation of high-frequency ''pseudo-modes", which reached
a maximum amplitude for a moderate field strength of $\approx600$~G. It was  suggested that
this may qualitatively explain the phenomenon of "acoustic halo" observed around sunspots
and active regions in the frequency range $5.5-7.5$~mHz (see, {\it e.g.}, \opencite{jain2002}).

There is no doubt that the inclination of the magnetic field also has a significant influence
on the oscillation properties of the resonant modes. This may be particularly important
in the sunspot penumbra, where the magnetic field is strong and highly inclined. The sunspot
penumbra represents a complicated mixture of almost horizontal magnetic fields, and also has
a circular structure with a radial dependence of the field properties. Therefore, the interpretation
of observations of penumbra oscillations in terms of the field properties is not straightforward.

 \begin{figure}
 \centerline{\includegraphics[scale=0.35]{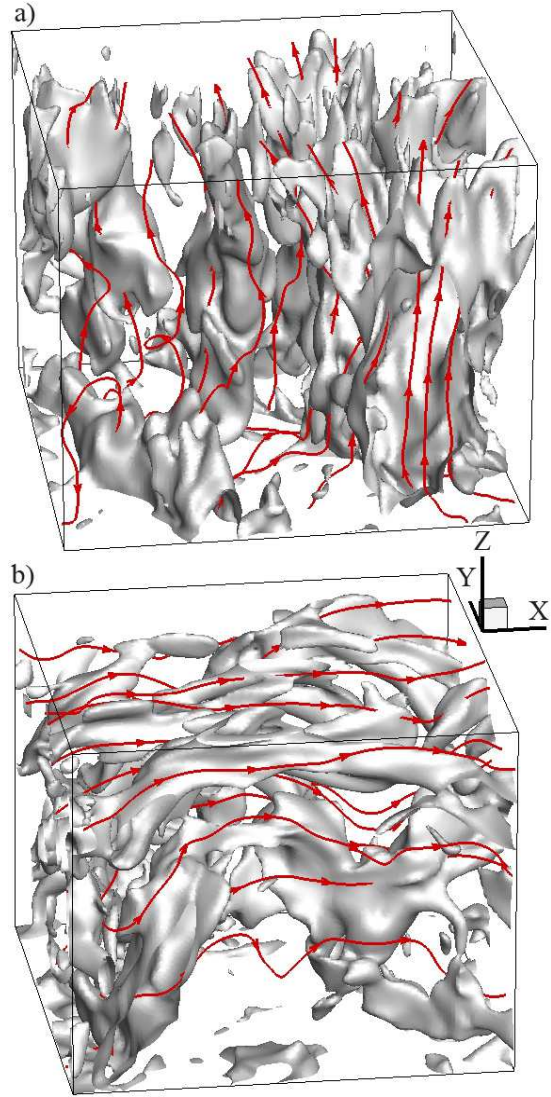}}
 \caption{Structure of the magnetic field magnitude ($B_0=600$~G) for different mean
 inclinations: a) vertical field ($\alpha=0^\circ$) and b) highly inclined field
 ($\alpha=85^\circ$). Isosurfaces correspond to 1100~G (a) and 1200~G (b) magnetic
 field strength. Red curves are magnetic field lines.}
 \label{3D-snapshots600}
 \end{figure}

In our simulations, we model a small area of a penumbra, where the mean magnetic field
strength is almost uniform. This simplifies the simulations and allows us to obtain
dependencies on the strength of the inclined field. For this initial investigation,
we used our simulations data \cite{kiti09a} previously obtained by using a radiative
MHD code {\it SolarBox} (Jacoutot {\it et al.}, 2008a; 2009), for the top 5~Mm-deep layer
of the convective zone and $0.5$~Mm layer of the low atmosphere with spatial resolution
50~km$\times$50~km$\times$42~km. Radiative transfer was calculated with a local thermodynamic equilibrium (LTE) approximation
using four-bin opacity distribution function. The ray-tracing transport calculations
implemented the Feautrier method for a 14 ray (two vertical, four horizontal, eight slanted)
angular quadrature. The code was tested \cite{jacoutot08a} by comparing with the results of
similar code \cite{nordlund2001,stein2001}, and also with some cases of higher-resolution
simulations (12.5~km$\times$12.5~km$\times$10.5~km). In the simulations the mean
magnetic field strength and inclination are maintained by the boundary conditions
\cite{kiti09a}. The simulation results have been used for studying the basic features
of the Evershed flow and the filamentary structures, including the ''sea-serpent"
behavior of magnetic field lines in the penumbra \cite{kiti10a}.

 \begin{figure}
 \centerline{\includegraphics[scale=0.35]{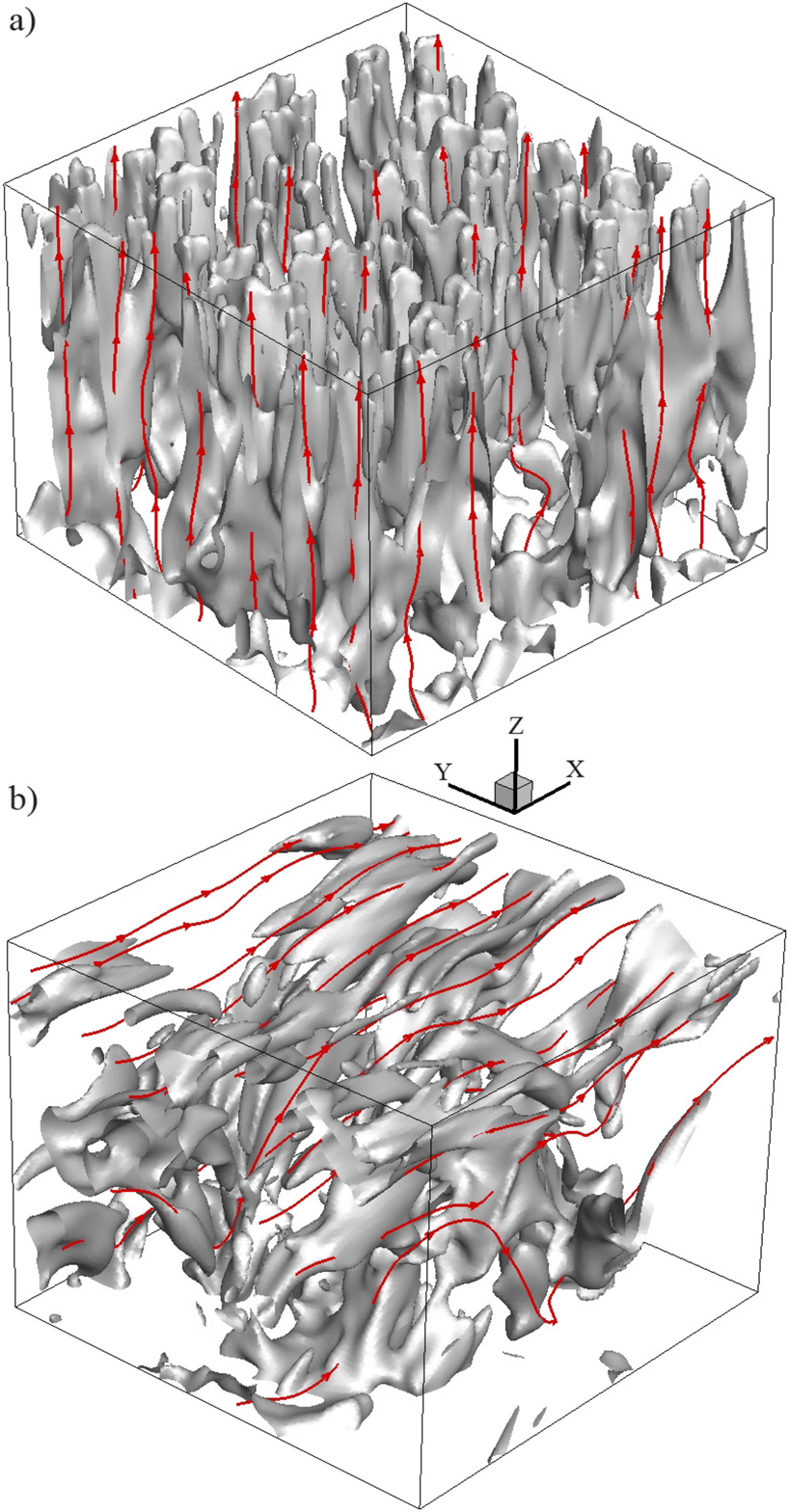}}
 \caption{Structure of the magnetic field magnitude ($B_0=1200$~G) for different mean
 inclinations: a) vertical field ($\alpha=0^\circ$) and b) highly inclined field
 ($\alpha=85^\circ$). Isosurfaces correspond to 1450~G (a) and 1700~G (b) magnetic
 field strength. Red curves are magnetic field lines.}
 \label{3D-snapshots}
 \end{figure}

\section{Magnetoconvection in Vertical and Highly Inclined Magnetic Field Regions}

 \begin{figure}[t]
 \centerline{\includegraphics[scale=1]{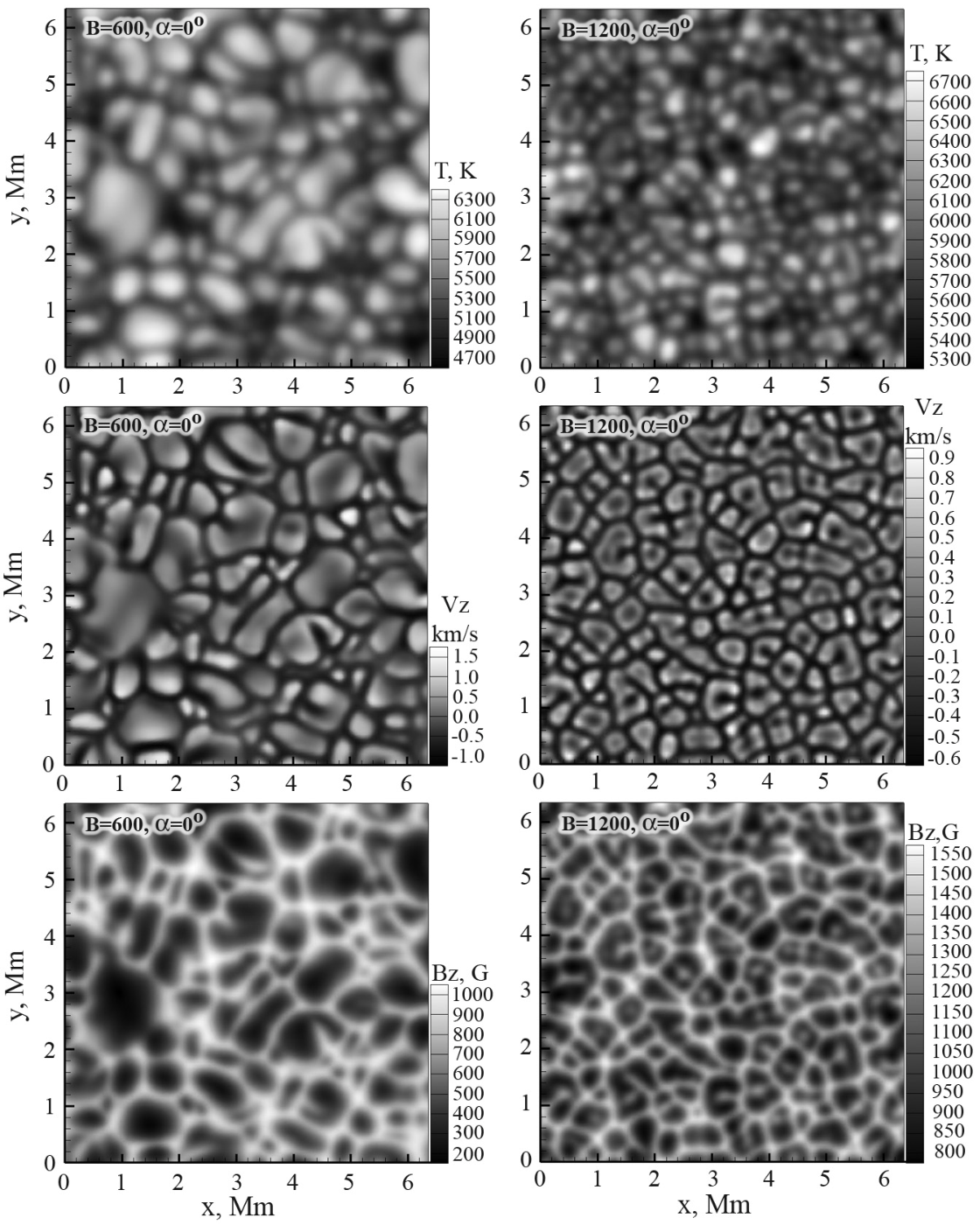}}
 \caption{Snapshots (from top to bottom) of temperature, {\it T}, vertical velocity, {\it Vz},
 and vertical magnetic field, {\it $B_z$}, for initial vertical field $B_{z0}=600$~G ($\alpha=0^\circ$,
 left column) and $B_{z0}=1200$~G (right column), at a constant depth corresponding to the photospheric level,
 approximately defined at the optical depth of unity.}\label{V-snapshots}
 \end{figure}

Magnetic fields have a strong influence on convective motions. For instance, in regions with
strong vertical magnetic field, such as the sunspot umbra, convection is largely suppressed.
The umbra convection is characterized by horizontally small and vertically long convective cells,
and slow flows \cite{schussler2006,bharti2010,kiti10b}. Such convective structures are observed
as umbral dots (see, {\it e.g.}, \opencite{ortiz2010}). The horizontal magnetic fields tend to stretch
convective granules along the field lines as shown in the simulations of the horizontal flux
emergence \cite{cheung08,stein2010}. The interaction of convective motions and magnetic field
highly depends on the field inclination (see Figures~\ref{3D-snapshots600} and~\ref{3D-snapshots}).
It is natural that the vertical magnetic field
structure represents vertical flux tubes for a strong field (Figure~\ref{3D-snapshots}a).
However, in the inclined field ($\alpha=85^\circ$) of the same strength, the magnetic field structure
is very different and forms arch-like structures below the surface and almost horizontal elongated
tubes at the surface (Figures~\ref{3D-snapshots600}b,~\ref{3D-snapshots}b).

The influence of the vertical field on convection strongly depends on the field strength.
The weak field (about 1, 10~G) almost does not affect the plasma dynamics. The field is concentrated
in the intergranular lanes. When a moderate, 100~G, vertical magnetic field is imposed in an initially
nonmagnetic convection layer, it may lead to spontaneous formation of a stable, pore-like, magnetic
structure \cite{kiti10c}. However, in strong magnetic field regions the behavior of convection is different.
Figure~\ref{V-snapshots} shows such changes (from left to right) of the temperature,
vertical velocity, and vertical magnetic field for $B_0=600$~G (left column) and $B_0=1200$~G
(right column). The simulations show the decreasing size of the convective granules, slower evolution,
suppression of convective flows, and magnetic field concentrations in the intergranular lines.
This is in agreement with previous results \cite{schussler2006,jacoutot08b}.

 \begin{figure}[t]
 \centerline{\includegraphics[scale=1]{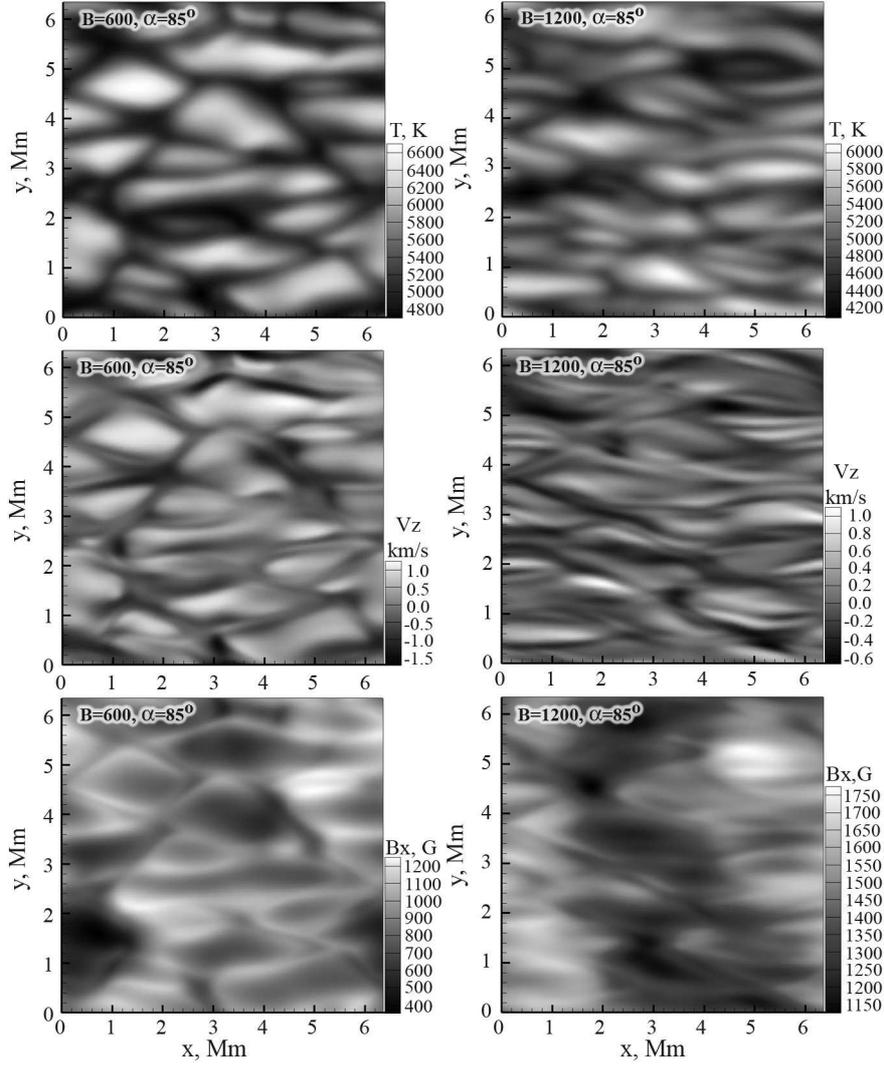}}
 \caption{Snapshots of temperature, {\it T}, vertical velocity, {\it Vz}, and horizontal component
 of magnetic field, {\it $B_x$}, for highly inclined ($\alpha=85^\circ$) initial field cases for
 $B_0=600$~G (left column) and $B_0=1200$~G (right column).}\label{In-snapshots}
 \end{figure}

The magnetic field inclination breaks the horizontal homogeneity of convection and leads to the
formation of a mean shear flow (Evershed flow) \cite{kiti09a}. The effect becomes stronger with
increasing field inclination and strength. Figure~\ref{In-snapshots} shows an example of
such changes for the 600~G (left column) and 1200~G (right column) initial magnetic field strengths,
and $85^\circ$ inclination.
 In this case, the convective cells are stretched in the direction of the magnetic field inclination.
 The degree of the shape deformation depends on both the field strength and the inclination angle.
 A weak inclination ($\approx30^\circ$) of a strong magnetic field (such as 1200~G) results in only
 a small stretching of granules along the magnetic field lines \cite{kiti10b}. When the magnetic field
 inclination increases, the magnetic effects on the granulation dynamics are much stronger.
 In particular, the high inclination leads to strong, horizontal mean flows resembling the Evershed
 effect \cite{kiti09a} and to formation of a filamentary structure in the form of strongly stretched
 convective cells, which become more narrow for the stronger field (see two bottom rows in
 Figure~\ref{In-snapshots}).

 Such changes of the solar magnetoconvection, depending on the inclination and strength of the magnetic
 field, can also be reflected in the oscillatory behavior.

 \begin{figure}[t]
 \centerline{\includegraphics[scale=0.38]{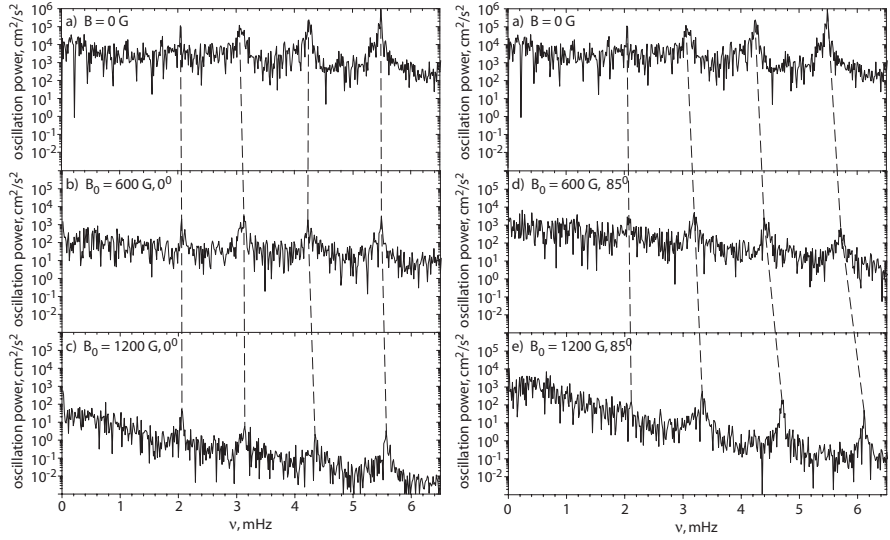}}
 \caption{Oscillation power spectra of the vertical velocity for different magnetic field strengths
 and inclinations: a) $B_0=0$~G; b) $B_0=600$~G, $\alpha=0^\circ$; c) $B_0=1200$~G, $\alpha=0^\circ$;
 d) $B_0=600$~G, $\alpha=85^\circ$; and e) $B_0=1200$~G, $\alpha=85^\circ$.}\label{pow-sp}
 \end{figure}

 \section{Frequency Shift of the Radial Oscillations}

We investigate effects of the magnetic field on the oscillations by calculating the power spectral
density for the vertical velocity fluctuations, averaged over the whole horizontal domain.
This corresponds to considering the radial oscillations or oscillations of a low angular degree
with the horizontal wavelength much larger than the size of the computational domain (6~Mm).
Because of the small domain of the simulations, we do not attempt to extract high-degree modes.
For calculating the power spectra, we used 20~hours of the simulated data sets of the vertical
velocity with a cadence of 30~seconds, after the magnetoconvection reached a stationary state.

Figures~\ref{pow-sp}~b-c illustrate the power spectrum changes with increasing vertical
magnetic field strength. Figures~\ref{pow-sp}~d, e show the power spectra for the highly inclined fields
of the same strengths. The two identical top panels (Figure~\ref{pow-sp}a) show the spectrum without
magnetic field, for comparison. In the frequency range of $0-6$ mHz, the power spectra have four
mode peaks, the frequencies of which are determined by the resonant conditions between the bottom
boundary conditions of our domain and the near-surface reflective layer. In this paper, we have not
considered the high-frequency spectrum because the top boundary condition was reflecting.

\begin{figure}[t]
 \centerline{\includegraphics[scale=0.4]{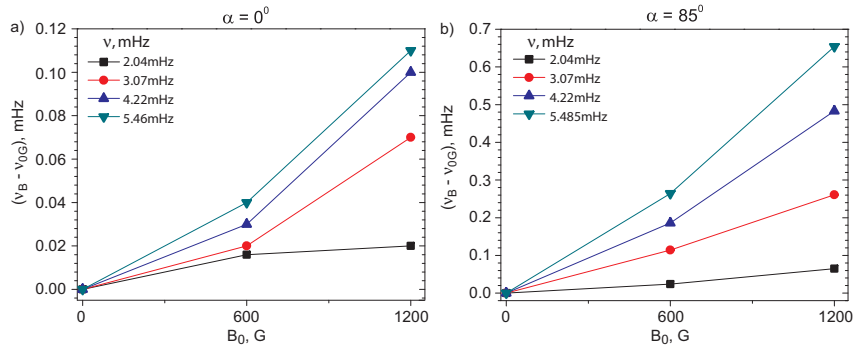}}
 \caption{Frequency shifts of the oscillation modes (indicated by different colors) relative
 to the case of nonmagnetic convection for the vertical ($\alpha=0^\circ$, panel a) and almost
 horizontal ($\alpha=85^\circ$, panel b) magnetic fields.}\label{shift}
 \end{figure}

 Comparison of the oscillation spectra in Figure~\ref{pow-sp} shows three main dependencies:
 {\it i}) suppression of the power by the magnetic field, {\it ii}) shift of the mode peaks
 to a higher frequency, which is greater for the stronger field, and {\it iii}) increase of the width
 of the resonant peaks with the magnetic field. In the vertical field, $B_{z0}=600$~G,
 the shift of the mode frequencies shows a trend to higher frequencies. The frequency shift increases
 in the stronger field, $B_{z0}=1200$~G. It is interesting that the power suppression is stronger
 for the vertical magnetic field than for the inclined field, but the frequency shift
 is significantly stronger for the inclined field than for the vertical field.  Also, the frequency
 dependence of the shift is stronger for the inclined magnetic field. The frequency shifts
 estimated from the positions of the peak maxima are shown in Figure~\ref{shift}. The dependence
 on the magnetic field strength seems to be nonlinear, in particular, for the two higher modes,
 but, obviously, this requires further investigation.

  \section{Conclusions}

We have used radiative MHD simulations of magnetoconvection in the upper convective boundary layer
and the low atmosphere of the Sun to investigate changes in the oscillation power spectrum
in the cases of vertical and highly inclined magnetic fields. The oscillations in these simulations
are naturally excited by the turbulent convection. We have investigated only the behavior of the radial
(or large-scale) oscillations.

The results show two basic effects of the magnetic fields: power suppression and shifts of the mode
peaks to higher frequencies. The power suppression is stronger for the vertical magnetic
field, but the frequency shift is stronger for the inclined magnetic field.
The stronger frequency shift in the inclined magnetic field can be understood in terms of physical properties
of fast MHD waves. The speed of these waves is higher when they travel across the magnetic field lines,
and this leads to higher modal frequencies. These results may
have important implications for local helioseismology analysis and for comparison with linear modeling of
oscillations in magnetic fields ({\it e.g.}, \opencite{parchevsky2009}). In particular, our results indicate
that the frequency shift measured in sunspot regions using the ring-diagram technique of local helioseismology
without discriminating the sunspot umbra and penumbra regions may come mostly from the penumbra because
the contributions of various parts of the analyzed area are weighted by the local oscillation power
of these parts.

{\bf Acknowledgement.} We thank the International Space Science Institute (Bern) for the opportunity to discuss these results at the international team meeting on solar magnetism.
%
%

\end{article}
\end{document}